\documentclass[%
 preprint,
 superscriptaddress,
 amsmath,amssymb,
]{revtex4-2}
\usepackage{graphicx}
\usepackage{dcolumn}
\usepackage{bm}
\usepackage{csquotes}
\usepackage[english]{babel}
\usepackage{siunitx}
\usepackage{booktabs}
\usepackage{tabularx}
\usepackage{adjustbox}
\usepackage{xcolor}
\usepackage{lineno}

\usepackage{braket}
\usepackage[normalem]{ulem}
\usepackage[caption=false, justification=centerlast]{subfig}
\usepackage{todonotes}

\newcommand{\rom}[1]{\uppercase\expandafter{\romannumeral #1\relax}}

\usepackage[%
colorlinks=true,
urlcolor=blue,
linkcolor=blue,
citecolor=blue
]{hyperref}

\usepackage{tabularx}

\newcommand{\UOL}{Department of Physics, The University of Liverpool, Liverpool, L69 3BX, United Kingdom}
\newcommand{\CI}{Cockcroft Institute, Warrington WA4 4AD, United Kingdom}

\newcommand{\UOM}{Department of Physics and Astronomy, The University of Manchester, Manchester M13 9PL, United Kingdom}

\newcommand{\ICMUV}{ICMUV, Universidad de Valencia, 46071 Valencia, Spain}

\newcommand{\FUHC}{Federal University of Health Sciences of Porto Alegre, Porto Alegre, RS, 90050-170, Brazil}

\newcommand{\GILPA}{Guangdong Institute of Laser Plasma Accelerator Technology, Guangzhou, China}
\newcommand{\IFIC}{Instituto de F\'isica Corpuscular (IFIC), Universitat de Val\`encia - Consejo Superior de Investigaciones Cient\'ificas, 46980 Paterna, Spain}

\begin{document}
	      

\title{Leaky surface plasmon-based wakefield acceleration in nanostructured carbon nanotubes  
}

\author{Bifeng Lei}
\email{bifeng.lei@liverpool.ac.uk}
\affiliation{\UOL}
\affiliation{\CI}
\author{Hao Zhang}
\affiliation{\UOL}
\affiliation{\CI}

\author{Cristian Bonţoiu}
\affiliation{\UOL}
\affiliation{\CI}

\author{Alexandre Bonatto}
\affiliation{\FUHC}

\author{Pablo Mart\'{i}n-Luna}
\affiliation{\IFIC}

\author{Bin Liu}
\affiliation{\GILPA}
\author{Javier Resta‐L\'{o}pez}
\affiliation{\ICMUV}
\author{Guoxing Xia}
\affiliation{\UOM}
\affiliation{\CI}
\author{Carsten Welsch}
\affiliation{\UOL}
\affiliation{\CI}

\date{\today}


\begin{abstract}

Metallic carbon nanotubes (CNTs) can provide ultra-dense, homogeneous plasma capable of sustaining resonant plasma waves—known as plasmons—with ultra-high field amplitudes. These waves can be efficiently driven by either high-intensity laser pulses or high-density relativistic charged particle beams.
In this study, we use numerical simulations to propose that electrons and positrons can be accelerated in wakefields generated by the leaky electromagnetic field of surface plasmons. These plasmons are excited when a high-intensity optical laser pulse propagates paraxially through a cylindrical vacuum channel structured within a CNT forest.
The wakefield is stably sustained by a non-evanescent longitudinal field with $\si{TV/m}$-level amplitudes. This mechanism differs significantly from the plasma wakefield generation in uniform gaseous plasmas.
Traveling at the speed of light in vacuum, with phase-matched focusing fields, the wakefield acceleration is highly efficient for both electron and positron beams.
We also examine two potential electron injection mechanisms: edge injection and self-injection. Both mechanisms are feasible with current laser facilities, paving the way for experimental realization.
Beyond presenting a promising pathway toward ultra-compact, high-energy solid-state plasma particle accelerators, this work also expands the potential of high-energy plasmonics.
\end{abstract}

\maketitle

\section{Introduction}


Laser plasma wakefield acceleration (LWFA) is a leading next-generation accelerator technology~\cite{Tajima:1979aa}. It enables the development of compact, portable systems (e.g., for medical and industrial applications) and large-scale scientific facilities, such as those advancing high-energy physics by extending the energy frontier~\cite{Tajima:2020aa}.
The exceptional performance of LWFAs arises from their ability to generate high acceleration gradients which depends on the plasma density $n_e$ as $E = m_e c \omega_p /e \simeq 9.6 \sqrt{n_e [10^{22} \si{cm^{-3}}]} ~[\si{TV/m}] \propto n_e^{1/2}$, where $n_e$ is the ambient plasma electron density, $m_e$ and $c$ are rest electron mass and speed of light in vacuum respectively, and $\omega_p = \sqrt{4\pi e^2 n_e/m_e}$ is the plasma frequency~\cite{Esarey:2009aa}. 
The recent progresses in LWFA are mainly based on the gaseous plasma where the density is typically in the range of $10^{16} - 10^{18}~\si{cm^{-3}}$. This density can, in principle, support an acceleration gradient of $1-100~\si{GV/m}$. For example, the experiments at BELLA have accelerated electrons to $10~\si{GeV}$ in $30~\si{cm}$ long plasma~\cite{Picksley:2024aa}, which indicates the acceleration gradient of $30~\si{GeV/m}$. This value scales with the theoretical prediction with the plasma density used in the experiment of $10^{17}~\si{cm^{-3}}$.
To maximize LWFA capabilities—whether for miniaturizing accelerators or achieving higher particle energies—we must prioritize the use of denser plasmas.
High-density gaseous plasma can be produced by high-pressure gas jets where the plasma density can reach near-critical values for optical lasers, approximately $10^{21}~\mathrm{cm}^{-3}$, and is intrinsically homogeneous~\cite{Henares:2019aa}. However, using such high-density gas targets for plasma wakefield excitation is unfeasible because optical laser pulses would be severely attenuated in the extended density gradients at the gas jet edges. Furthermore, at densities approaching the critical density, optical lasers cannot efficiently propagate due to strong reflection and absorption, preventing effective wakefield formation.

Naturally, super-dense plasma exists in solids.  
Over other materials including amorphous or polycrystalline solids, crystals have long-range ordered atomic structure (periodic lattice) and then provide several advantages to be the ideal candidates for the high-density plasma wakefield excitation. 
Their unique crystalline structure and ionization efficiency ensure that crystals can provide higher plasma density of intrinsically homogeneous distribution, up to $10^{24}~\si{cm}^{-3}$ typically or even higher to $10^{28}~\si{cm}^{-3}$ for ultra-dense materials, e.g. diamond anvil cells. 
As a result, coherent behaviors, such as, resonant plasma waves (so-called plasmons), can dominate~\cite{Pines1953aco}.
These plasmons can be excited either within the bulk of solid plasma or at the interface between a metal and a surrounding medium and are referred as bulk plasmon or surface plasmon (SP), respectively. 
Such solid plasma can, in principle, support ultra-high acceleration gradient, e.g. up to $\si{PeV/m}$.
However, natural crystal-based channel wakefield accelerator is unrealizable at current stage due to the limitation of angstrom-size channels~\cite{Tajima1987cry, Ammosov:1986aa}. 

It was suggested that carbon nanotubes (CNTs) may provide a solution to increase the channel size to $\si{nm}$ scale by compromising the acceleration gradient down to $\si{TeV/m}$~\cite{Shin:2019aa, Zhang:2016aa}. 
CNTs are graphene-based synthetic nanostructures formed by rolling single or multiple graphene sheets into cylindrical shapes.
With inner diameters typically spanning a few nanometers and lengths extending up to millimeters, CNTs exhibit exceptional mechanical, thermodynamic, and electronic properties~\cite{Eatemadi:2014aa, Rathinavel:2021aa}.
Depending on their chirality, CNTs can behave as metallic conductors (armchair type)~\cite{Fischer:1999aa, Ando:2009aa} with the potential to support an electron number density up to the order of $10^{24}~\si{cm^{-3}}$ in individual CNTs. Hereafter, we refer to this type of metallic CNT simply as CNT for clarity.
However, a single CNT currently cannot support efficient particle acceleration due to the mismatch between the $\si{\mu m}$-scale dimensions of available photon or particle beams and the natural nanoscale dimensions of CNTs.
As a result, direct experimental demonstrations are still lacking, as achieving resonant coupling between CNT structures and high-energy beams remains a challenge. 

\begin{figure*}
	\centering
	\includegraphics[width=0.9\textwidth]{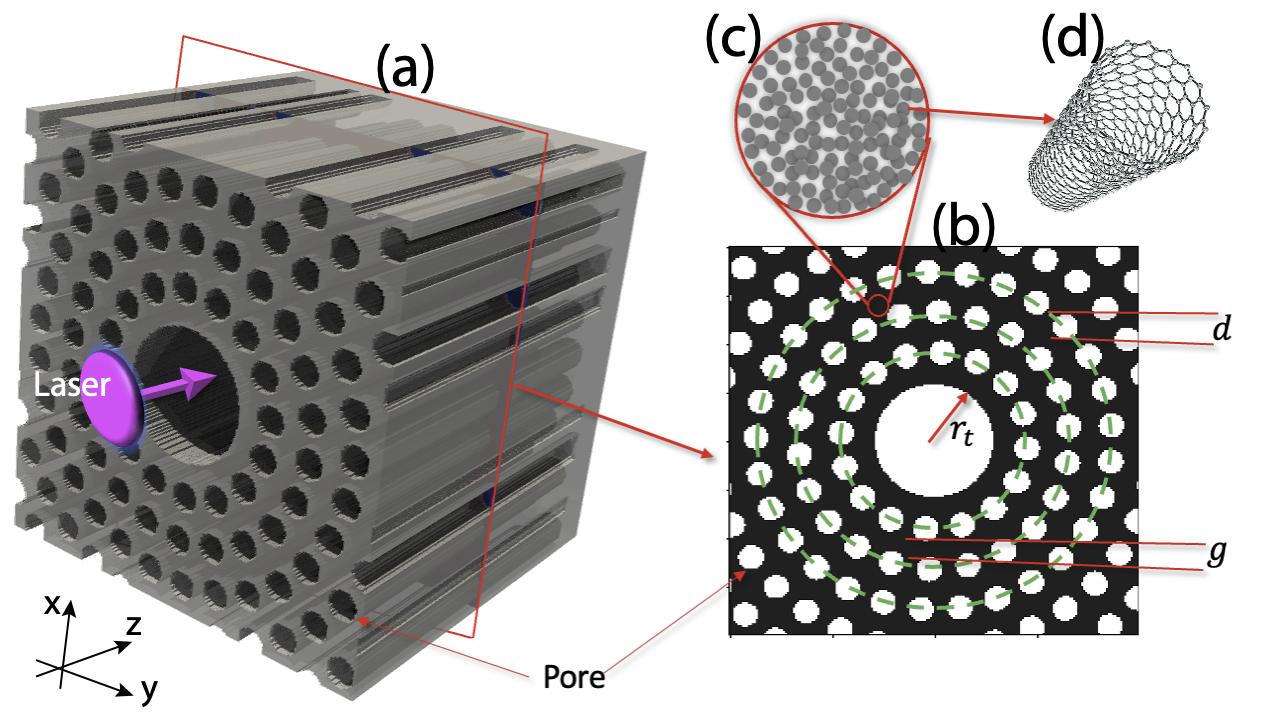}
	\caption{Schematic representation of (a) a nanostructured CNT forest target in a porous configuration and (b) its transverse layout, showing key parameters such as pore diameter $d$, central channel radius $r_t$, and virtual layer gap $g$.
(c) A zoomed-in view of the transverse structure of the CNT forest, where each grey dot represents an individual single-wall or multi-wall CNT, as further detailed in (d).
The laser pulse (represented by the pink spot in (a)) propagates along the central channel in the $z$ direction.}
	\label{fig:target}
\end{figure*}

Modern nanotechnologies enable the fabrication of structured CNTs in dense forest form, an organised array of vertically aligned CNTs~\cite{Li:1999aa, Sarasini:2022aa}. 
Unlike unstructured CNTs, the CNT forests offer significant flexibility in terms of geometry, density, and surface structure~\cite{Futaba:2006aa, Iijima:1991aa, Yang:2018aa}.
This allows the CNT targets to be designed with the $\si{\mu m}$-scale channels to guide the optical laser pulse through for a long distance as illustrated in Fig.~\ref{fig:target}. 
These structures are typically synthesised using advanced techniques like chemical vapour deposition (CVD)~\cite{Robertson:2012aa, Sugime:2021aa} or anodic aluminium oxide (AAO) template method~\cite{Hou:2012aa}. The AAO technique, in particular, offers exceptional tunability by enabling precise control over the pore structure of the template and geometry of the channel, e.g. cylindrical or square with optimized parameters.
The length of the CNT forest can extend up to $10\text{–}20~\si{mm}$~\cite{Sugime:2021aa}.
The density can be precisely controlled by adjusting parameters such as the packing density or the volume fraction of CNTs~\cite{Yu:2009aa}, achieving typical densities in the range of $n_e \sim 10^{19 - 23}~\si{cm^{-3}}$.
These unique features promise the nanostructured CNTs as an outstanding candidate for the laser-driven solid plasma wakefield accelerators.

In this paper, we propose laser-driven plasma wakefield excitation for both electron and positron acceleration in a nanostructured CNT forest channel.
It is shown that the unique characteristics of this kind of plasma wakefield—including its excitation mechanisms and field dynamics—are closely linked to SPs excitation and are systematically investigated through theoretical modelling and numerical simulations.
Our results show that, at moderate intensity, the CNT electrons remain confined to the inner walls of the forest channel. The electromagnetic (EM) field of resulting SPs leaks into the vacuum channel, forming a wakefield that may be used to accelerate both electrons and positrons. As the laser intensity increases, the electrons on the CNT surfaces undergo stronger oscillations, with some electrons escaping into the vacuum channel, where they become trapped and accelerated by the wakefield. In this regime, the wakefield amplitudes reach several teravolts per metre ($\si{TV/m}$). At even higher intensities, CNT electrons can freely traverse the surface, forming a plasma bubble of the high amplitude electric field, similar to those observed in gaseous plasma~\cite{Lu:2006aa}.
Two injection mechanisms are observed: edge injection and self-injection. In edge injection, a well-defined electron beam with a narrow energy spectrum (sub-$\si{\mu m}$ beam size, $\si{fs}$ duration and $\si{nC}$ charge) is trapped as the laser pulse enters the forest channel. In contrast, self-injection is a continuous process in which electrons emitted from the CNT walls enter the central vacuum region, leading to the formation of an electron beam with a broader energy spectrum and $\si{nC}$ charge.
This work provides valuable insights into the wakefield excitation in CNTs, thereby advancing the prospects for compact, highly efficient particle accelerators with experimental feasibility.

\section{Surface plasmon and wakefield generation in structured CNT forest channels}
A high-intensity laser pulse propagating along a plasma-vacuum boundary excites the confined surface waves which can be treated as the polarisation of the plasma in response to the laser field~\cite{Dragila:1988aa}.
The dispersion of SP with relativistic effects is given by 
\begin{equation}
	1+\varepsilon \sqrt{\frac{n^2-1}{n^2-\varepsilon}} = 0 \mathrm{,}
	\label{eq:dispersion}
\end{equation}
where $\varepsilon=1-\omega_p^2/ \gamma_e \omega_{sp}^2$ and $\gamma_e$ is the Lorentz factor of the electron. $n=c k_z/\omega_{sp}=c/v_{\text{ph}}$ with $\omega_{sp}$ and $v_{\text{ph}}$ the frequency and phase velocity of the SP, respectively. $k_z$ is the longitudinal component of the wavevector. $\varepsilon = -1$ in resonance excitation with the frequency $\omega_{sp}=\omega_p/\sqrt{2 \gamma_e}$~\cite{Ritchie1957pla}.  Eq.~\eqref{eq:dispersion} can only be satisfied if $n>1$. $n>1$ indicates that the phase velocity $v_{ph}$ is less than $c$, leading to the dephasing problem.
The phase velocity with resonance condition is given as 
\begin{equation}
	v_{\text{ph}} = c \sqrt{(\varepsilon-2)/(\varepsilon-1)} = c \sqrt{1 - \frac{1}{\alpha}} < c \mathrm{,}
	\label{eq:vph}
\end{equation}
where $\alpha = 2 \gamma_e -1$. In the non-relativistic case $\gamma_e \simeq 1$, $v_{\text{ph}} \simeq 0$, indicating that SP is a standing wave with a dominant electrostatic component~\cite{Dragila:1988aa} and is unsuitable for relativistic particle acceleration.
In the ultrarelativistic case $\gamma_e \gg 1$, for example, driven by a high-intensity laser pulse, $v_{\text{ph}}$ approaches $c$, which allows the acceleration of relativistic particles.  SP excitation also depends on the surface geometry.

\subsection{On flat surface}

\begin{figure*}
	\centering
	\includegraphics[width=0.9\textwidth]{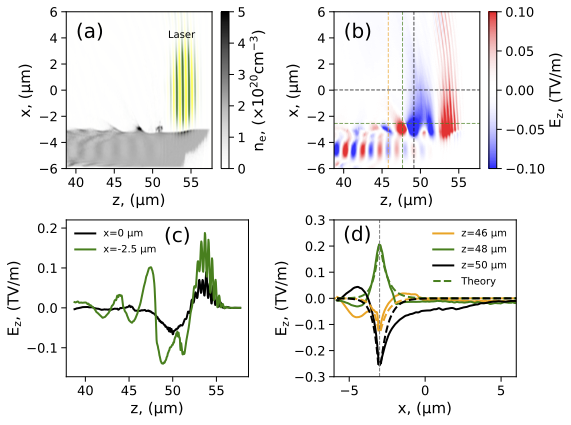}
	\caption{PIC results: SPs excitation on the flat CNT surface. (a) Density distribution of wall electrons (grey colourmap) and laser field (green-yellow colourmap). 
	(b) Longitudinal acceleration field $E_z$.
	(c) Line plots of $E_z$ along $x=0~\si{\mu m}$ (black) and $x=-2.5~\si{\mu m}$ (green), indicated by the horizontal dashed lines in (b), respectively.
	(d) Line plots of $E_z$ along $z=46~\si{\mu m}$ (orange), $z=48~\si{\mu m}$ (green) and $z=49~\si{\mu m}$ (black), indicated by the vertical dashed lines in (b) respectively. The vertical grey line indicates the position of the interface between the CNT wall and the vacuum. The corresponding theoretical estimations are shown by the dashed lines, respectively.}
	\label{fig:flat_SPs}
\end{figure*}

We start from the SP excitation on a solid flat surface in order to present the results with the cylindrical geometry.
A typical SP excitation on the flat surface is shown  Fig.~\ref{fig:flat_SPs} from a full 3D Particle-In-Cell (PIC) simulation using the code WarpX~\cite{Fedeli:2022aa}.
We can assume the surface (grey colour) is made of CNT forest and is initially modelled with neutral uniformly distributed Carbon atoms with the inization energies modified with the CNT properties, including weak $\pi$-bond and C-C bond ($\sigma$-bond) energies. The thermionic emission of electrons is negligible by assuming the room temperature. Field ionisation is implemented using the Ammosov-Delone-Krainov (ADK) method ~\cite{Ammosov:1986aa}.
The surface is positioned at $x = -3.0~\si{\mu m}$ and extends infinitely in $x\leq -3.0~\si{\mu m}$ and $y$ direction, and in range $0~\si{\mu m}\leq z \leq 60~\si{\mu m}$. 
Two vacuum sections of $20~\si{\mu m}$ and $10~\si{\mu m}$ are placed longitudinally at the head and tail of the target, respectively, for laser initialisation and electron beam diagnostics.
A $5~\si{TW}$ Gaussian laser pulse of wavelength $\lambda_L=0.8~\si{\mu m}$ is linearly polarised in the $y$ direction and propagates in the $z$ direction.
It is focused at $z=0~\si{\mu m}$ with root-mean-square (RMS) waist $w_0=3.0~\si{\mu m}$ and duration $\sigma_{\tau}=6.0~\si{fs}$. These parameters give a normalised laser strength $a_0=4.0$ or intensity $I=3.4\times 10^{19}~\si{W/cm^2}$, with the field intensity of $1.46\times 10^{19}~\si{W/cm^2}$ on the surface at $x = -3.0~\si{\mu m}$ which drives the relativistic electron dynamics.
The dimensions of the moving window are $12~\si{\mu m} \times 12~\si{\mu m} \times 20~\si{\mu m}$, comprising $384\times 384 \times 512$ cells in the $x$, $y$, and $z$ directions, respectively. Each cell contains 8 macro particles, which is sufficient to solve the electron dynamics effectively. For particle and field updates, the simulations utilise the Boris pusher~\cite{boris1970re} and the CKC Maxwell solver~\cite{Cowan2013ge}.

This physical phenomenon can be theoretically studied by solving Maxwell's equations with appropriate boundary conditions at the surface~\cite{Pitarke:2007aa}. The longitudinal electric field can be expressed as
\begin{equation}
	E_z = E_{z0} e^{-k_{x} x} e^{i(k_{z}z - \omega_{sp} t)} \mathrm{,}
	\label{eq:SP_flat}
\end{equation}
where $E_{z0}$ is the amplitude and $k_x = \omega_{sp} \sqrt{n^2-1}/c$.
It is shown in Eq.~\eqref{eq:SP_flat} that the EM field decays exponentially away from the surface which can be observed from PIC simulations as shown in Fig.~\ref{fig:flat_SPs}(c).
Both longitudinal and transverse SP fields with an amplitude of $0.1~\si{TV/m}$ can leak into a small vacuum region near the surface and decay exponentially, as shown in Fig.~\ref{fig:flat_SPs}(b) to (d). 
The resonance wavelength is theoretically calculated to be $\lambda_{sp,Theo} = 3.24~\si{\mu m}$, which agrees well with the simulation result $\lambda_{sp,PIC} \simeq 3.3~\si{\mu m}$.
SP decays slower in vacuum than in CNTs, since $k_{x,\text{vacuum}} = \omega_{sp}\sqrt{n^2-1}/c < k_{x,\text{CNTs}}=\omega_{sp}\sqrt{n^2-\varepsilon} /c$~\cite{Dragila:1988aa}.
The longitudinal component $E_z$ along the $x$-direction is plotted at three different positions by solid lines in Fig.\ref{fig:flat_SPs}(d) and shows good agreement with the theoretical estimates indicated by the dashed lines. 
These agreements benchmark the validation of PIC simulations for SP dynamics.
The observed deviations, e.g. at $z = 50~\si{\mu m}$ as shown by the solid and dashed black lines, are due to the influence of the laser field.
With the evanescent nature of the leaky EM field from the flat surface, the accelerating volume is limited to a small region, e.g. $|\Delta x|<1/k_{sp}~\sim 0.2~\si{\mu m}$ for our case, making the acceleration highly sensitive to the angular distribution of both the driving and witness beams. Consequently, the total charge and energy gain of the witness beam are limited. Therefore, it is not suitable for high-energy particle acceleration.

\subsection{On cylindrical surface}
\begin{figure}
	\includegraphics[width=0.49\textwidth]{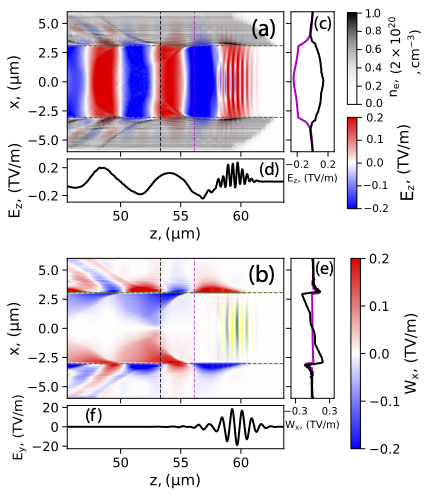}
	\caption{PIC results after laser pulse propagating $z=60~\si{\mu m}$ inside the CNT forest channel: (a) Conduction electron density distribution $n_e$ (grey colourmap), longitudinal acceleration field $E_z$ (blue-red colourmap) in $xz$-plane sliced at $y=0~\si{\mu m}$. 
		(b) Transverse field $W_x=E_x - c B_y$ in $xz$-plane sliced at $y=0~\si{\mu m}$. The green-yellow colourmap shows the laser field.
	(c) Transverse slice of $E_z$ along the dashed pink and black lines in (a), respectively. 
	(d) On-axis ($x=0~\si{\mu m}$) plot of $E_z$.
	(e) Transverse slice of $W_x$ along the dashed pink and black lines in (b), respectively. 
	(f) On-axis ($x=0~\si{\mu m}$) plot of $E_y$.
	The horizontal green dashed lines in (a) and (b) indicate the positions of the inner surface of the forest.}
	\label{fig:SPs_field}
\end{figure}

The structure of the leaky EM field structure becomes different on a cylindrical surface which can be nanostructured into the CNT forest as shown in Fig.~\ref{fig:target}, where the radius of the surface is $r_t$. 
With bulk dimensions significantly larger than those of a single CNT, the inter-structural interactions between single CNT can be approximated as effectively uniform~\cite{Bonatto:2023aa}. 
Without couplings between SP modes of different azimuthal indices, the leaky field can remain monochromatic with the resonant frequency, as seen in Fig.~\ref{fig:SPs_field}(a) and (b).
The PIC simulations are done with a $5~\si{TW}$ Gaussian laser pulse same as that used in Fig.~\ref{fig:flat_SPs} passing through the vacuum channel of a nanostructured CNT forest target arranged in a porous pattern as seen in Fig.~\ref{fig:target}. 
The target is designed with a cylindrical vacuum channel of radius $r_t=3.0~\si{\mu m}$.
Details of the target configuration are summarised in Table.\ref{tab:target_paras}.
The numerical parameters are also the same as those in Fig.~\ref{fig:flat_SPs}.

\begin{table}[ht]
\caption{Parameters of the porous CNT target used in PIC simulations. }
\centering
\begin{tabular}{p{5cm}p{3cm}}
\hline
Layer gap, $g$ & $50~\si{nm}$  \\ 
Pore diameter, $d$ & $100~\si{nm}$ \\ 
Forest channel radius, $r_t$ & $3~\si{\mu m}$ \\ 
Filling factor, $f_f$ & $0.65$ \\  
Number of layers, $N$ & $20$  \\
Wall density, $n_e$ & $2\times 10^{20}~\si{cm}^{-3}$ \\ \hline
\end{tabular}	
\label{tab:target_paras}
\end{table}

SP excitation and the leaky field in the vacuum channel are shown in Fig.~\ref{fig:SPs_field}. 
The wall electrons oscillate collectively driven by the ponderomotive scattering with the laser pulse and are confined to the surface due to the transverse ionic potential, as shown in Fig.~\ref{fig:SPs_field}(a), where the laser pulse has propagated $60~\si{\mu m}$ within the forest channel. 
A microscale monochromatic leaky field inside the vacuum channel is excited. The wavelength obtained from PIC, $\lambda_{sp, \text{PIC}}=5.0~\si{\mu m}$, agrees well with the theoretical prediction with relativistic effect, $\lambda_{sp, \text{Theo.}}=2 \pi c/\omega_{sp}=4.8~\si{\mu m}$.
The longitudinal component of the leaky field remains quasi-uniform across the vacuum channel with an amplitude of $0.2~\si{TV/m}$, providing periodic accelerating phases for both negatively and positively charged particles, as shown in Fig.~\ref{fig:SPs_field} (a), (c), and (d), which is advantageous for maintaining beam quality.
The transverse component of the field, $W_x=E_x - c B_y$, supplies a focusing force for positrons, phase-matched with the corresponding acceleration phase, as shown in Fig.~\ref{fig:SPs_field}(b) and by the black line in (e). This field reaches an amplitude of $0.3~\si{TV/m}$ on the surface and decays transversely.
In the phase for electron acceleration, the transverse field vanishes in the vacuum region, instead providing a confining force on the inner surface, as shown by the green line in Fig.~\ref{fig:SPs_field}(e). 
Consequently, electrons can also be confined within the vacuum channel and accelerated.
The laser pulse propagates through the vacuum channel without any noticeable deformation, as shown in Fig.~\ref{fig:SPs_field}(b) and (f). This demonstrates that the scheme is dynamically stable and can be effectively scaled for longer targets, limited only by the depletion of the laser energy.

Theoretically, we can understand harmonic wakefield dynamics with the assumption that a channel radius $r_t$ is large, as $k_{sp} r_t \gg 1$. In this case, the eigenfrequencies of all SP modes become less sensitive to surface curvature and approach the resonant frequency~\cite{Schmeits1989sur, Warmack1984sur}.
Since the length of the target is significantly larger than the channel radius ($L \gg r_t$), the electric potential of leaky SP modes in a cylindrical vacuum channel can be derived by solving the Laplace equation in cylindrical coordinates $(r, \phi, z)$ as follows~\cite{Schmeits:1988aa}:
\begin{equation}
	\varphi_m (r, \phi, z,t) = \frac{\varphi_A }{I_{m}(k_{sp} r_t)} I_{m}(k_{sp} r) e^{i m \phi} e^{i (k_{sp} z -\omega_{sp} t)} \mathrm{,}
	\label{eq:phi_SP}
\end{equation}
where $\varphi_A$ is the amplitude and $m$ is an integer index. The electric field can be written as $\bm{E}= - \nabla \varphi$.  
It can be seen that only the axial mode  $m = 0$  does not decay to zero at any point within the vacuum channel, while the other modes decay rapidly and are zero at  $r = 0$. This rapid decay renders the other modes unsuitable for efficient acceleration.
The longitudinal component of the leaky field is given by
\begin{equation}
	E_z = -\partial \varphi_0 / \partial z = - i k_{sp}\frac{ \varphi_A}{I_{0}(k_{sp} r_t)} I_{0}(k_{sp} r) e^{i (k_{sp} z -\omega_{sp} t)} \mathrm{,}
	\label{eq:leaky_Ez}
\end{equation}
which remains non-zero inside the vacuum channel. 
The mode $m=0$ corresponds to a quasi-uniform leaky accelerating field with cylindrical symmetry as presented in Fig.~\ref{fig:SPs_field} (a). 
The transverse component is given by
\begin{equation}
	W_r = -\partial \varphi_0 / \partial r = -k_{sp}\frac{\varphi_A }{I_{0}(k_{sp} r_t)} I_{1} (k_{sp} r) e^{i (k_{sp} z -\omega_{sp} t)} \mathrm{,}
	\label{eq:leaky_Er}
\end{equation}
which decays to zero and exists only in a narrow region near the surface, in agreement with the simulation results shown in Fig.~\ref{fig:SPs_field}(b) and \ref{fig:self_injection}(b).
From Eqs.~\eqref{eq:leaky_Ez} and \eqref{eq:leaky_Er}, it can be seen that there is a $\pi/2$ phase region where either negatively or positively charged particles can experience both acceleration and focusing forces. 
Therefore, the leaky field of the fundamental mode, $m=0$, is suitable for accelerating both electrons and positrons.

\begin{figure}
\centering
	\includegraphics[width=0.4\textwidth]{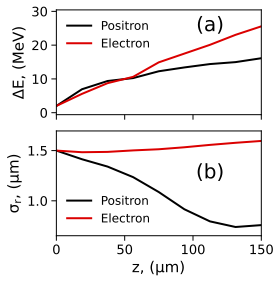}
	\caption{PIC Results:
(a) Evolution of the energy gain of witness positron (black) and electron (red) beams at the acceleration phase indicated by the vertical dashed lines in Fig.~\ref{fig:SPs_field}(a).
(b)  Evolution of radial beam size $\sigma_r = \sqrt{\sigma_x^2 + \sigma_y^2}$, where $\sigma_{x,y}$ denotes the transverse RMS sizes of witness positron (black) and electron (red) beams in $x$ and $y$ directions, respectively. }
	\label{fig:leaky_beam_acc}
\end{figure}

To be successfully accelerated, a relativistic electron or positron beam should be injected into the correct phase.
We now put either an electron or positron beam in the corresponding accelerating phase, indicated by vertical dashed pink and black lines in Fig.~\ref{fig:SPs_field}(a) and (b), respectively.  Both of the beams have the same parameters except for the negative or positive charge, as summarised in Table~\ref{tab:witness_beam}.

\begin{table}[ht]
\caption{Parameters of the witness positron and electron beams used in simulations. }
\centering
\begin{tabular}{p{5cm}p{3cm}}
\hline
Beam profile & Gaussian  \\ 
Charge, $Q_{e^{+}, e^{-}}$ & $\pm 1~\si{pC}$ \\ 
Beam energy, $\cal{E}$ & $200~\si{MeV}$ \\ 
Beam RMS size, $\sigma_r$ & $1.5~\si{\micro m}$ \\  
Beam RMS duration, $\sigma_{\tau}$ & $1.6~\si{fs}$ \\ 
Energy spread & $1.0\%$ \\ \hline
\end{tabular}	
\label{tab:witness_beam}
\end{table} 

The evolution of energy gain and transverse beam size is shown in Fig.~\ref{fig:leaky_beam_acc} (a) and (b), respectively. All the electrons and positrons inside the witness beam can be preserved and accelerated which indicates that the transverse component of the leaky field can sufficiently confine both beams. 
With the positron beam, after $z=150~\si{\mu m}$ propagation, the beam gains energy $\Delta E =19~\si{MeV}$ energy which indicates the mean energy gain rate as $R_E\approx 0.13~\si{TeV/m}$.
The beam size is reduced by half until being equilibrated at $z=130~\si{\mu m}$, where the focusing field is in balance with the self-space-charge field of the witness beam. 
For the witness electron beam, the energy gain is $\Delta E =25~\si{MeV}$, which indicates the mean energy gain rate as $R_E \approx 0.17~\si{TeV/m}$.
The radial beam size is preserved with a small increase as the transverse component of the leaky field at the acceleration phase for the electron beam is weaker than that for the positron beam as shown in Fig.~\ref{fig:SPs_field}(b) and (e).
Additionally, the quasi-linear energy gain indicates that the phase slippage between the witness beam and leaky field is slow and then enables the extensible acceleration for high-power laser pulse.

\section{Electrons injection}
\label{sec:injection}

In experiments, a positron beam must be externally injected into the appropriate phase of the leaky field to achieve acceleration, as the CNT target does not inherently generate positrons.
Currently, available positron sources can produce positron beams with energies at the $\si{MeV}$-level, making this approach feasible~\cite{Chaikovska:2022aa}.

For electron acceleration, it is more feasible that the wall electrons can be self-injected into the correct phase of the leaky field if these electrons can meet three criteria: firstly, they must be emitted freely from the inner surface into the vacuum channel; secondly, they must be relativistic to catch up with the phase velocity of the leaky field; and thirdly, they must be confined within the channel.  
The first condition is crucial, which requires the wall electrons to attain a sufficient transverse momentum to overcome the confinement on the surface. 
Here, we show two possible mechanisms that could efficiently inject the wall electrons into the correct phase for acceleration.

\subsection{Edge injection}
\begin{figure*}
\includegraphics[width=0.9\textwidth]{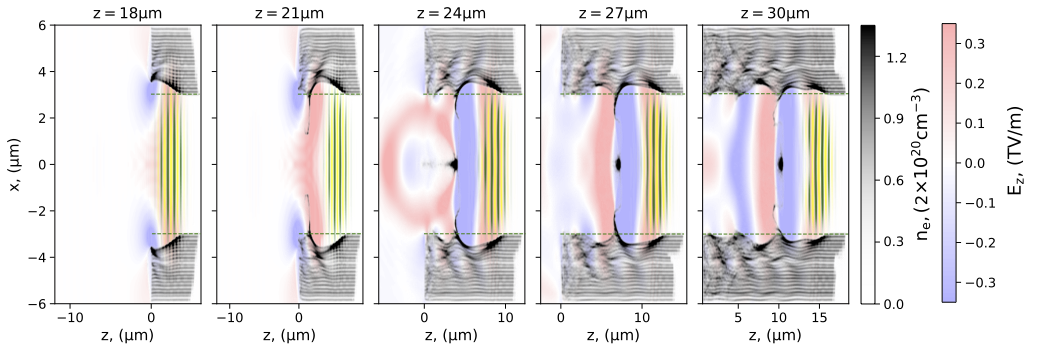}
\center
\caption{PIC results: Evolution of the electron density (grey colourmap), longitudinal electric field (blue-red colourmap), and laser field (green-yellow colourmap) at five propagation positions:  $z = 18, 21, 24, 27$ and $30~\si{\mu m}$. The laser strength is  $a_0 = 5.0$, with other parameters the same as those in Fig.~\ref{fig:SPs_field}. 
The horizontal green dashed lines indicate the initial positions of the inner surface of the forest.}
\label{fig:SPs_injection}
\end{figure*}

With a low-intensity laser pulse, wall electrons gain relatively low momentum and are unable to overcome the two-dimensional (2D) transverse confinement on the channel surface, preventing them from freely crossing the inner vacuum surface.
However, the situation changes during the laser enters the forest channel. Here, the vertical edge provides an additional dimension of confinement in the longitudinal direction, which compresses the streaming electrons, as shown in Fig.~\ref{fig:SPs_injection}.
This compression allows some electrons to gain additional momentum which allows them to overcome the transverse confinement and emit into the vacuum channel. Since the SPs are localised waves, the emitted electrons are naturally in the acceleration phase along the surface and are simultaneously accelerated longitudinally while drifting transversely through the vacuum channel.
These electrons collide at the channel centre and become trapped in the leaky field once they acquire sufficient longitudinal momentum.
This method is analogous to catapult injection at the edge of stacked graphene layers~\cite{Bontoiu:2023aa}.
PIC simulations reveal that after $100~\si{\mu m}$ of propagation, a $42~\si{pC}$ electron beam with a duration of $3.3~\si{fs}$ is accelerated to $35~\si{MeV}$ with a  $1\%$  energy spread, corresponding to a mean acceleration gradient of $0.35~\si{TeV/m}$. The beam transverse size is  $\sigma_r = 0.2~\si{\mu m}$.
Such an ultrashort, high-quality beam holds unprecedented potential for advanced applications in high-field science and beyond~\cite{Morimoto:2023aa}.

\subsection{Self injection}

\begin{figure*}
\centering
	\includegraphics[width=0.9\textwidth]{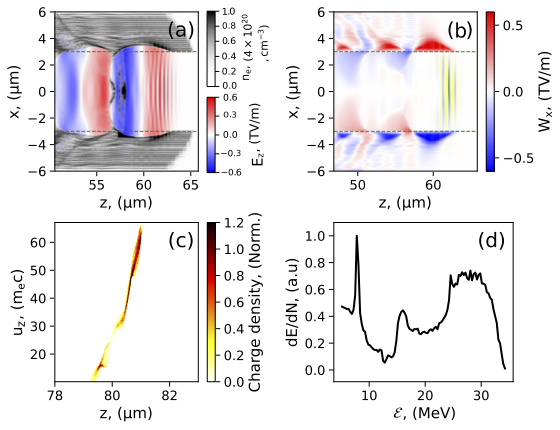}
	\caption{PIC results: 
	(a) Electron density distribution ($n_e$, grey colourmap), and longitudinal electric field ($E_z$, blue-red colourmap).
	(b) Transverse field, $W_x$ (blue-red colourmap) and laser field (yellow-green colourmap). 
	(c) Phase space of the accelerated electron beam after exiting the $80~\si{\mu m}$ long forest target.
	(d) Energy spectrum corresponding to (c).
	All the laser and target parameters are the same as those used for Fig.~\ref{fig:SPs_field} except the laser strength which is increased to $a_0=10$. The horizontal green dashed lines in (a) and (b) indicate the initial positions of the inner surface of the forest.}
	\label{fig:self_injection}
\end{figure*}

With increased laser strength, electron oscillations on the surface intensify, leading to a significant enhancement of the SP amplitude, as demonstrated in Fig.~\ref{fig:self_injection}(a), where a laser pulse with $a_0=10$ or peak power $P=30~\si{TW}$ is used.
This enhancement enables some wall electrons to gain sufficient transverse momentum to cross the inner surface and penetrate deeply into the vacuum region.
These electrons naturally traverse the forest surface through the accelerating phase of the SP and align with the accelerating phase of the leaky field in the vacuum channel. The leaky field, with a magnitude as high as $0.6~\si{TV/m}$, facilitates the rapid longitudinal acceleration of these transiting electrons, allowing them to catch up with the leaky field within just a few $\si{\mu m}$.
The transverse field of the SP acts as a confining force along the inner surface, preventing the transited electrons from escaping back into the forest wall, as illustrated in Fig.~\ref{fig:self_injection}(b). 
As a result, a substantial number of electrons are trapped and continue to be efficiently accelerated.

A $204~\si{pC}$ electron beam is self-injected and accelerated to $35~\si{MeV}$ over an $80~\si{\mu m}$ long-distance, corresponding to a peak acceleration gradient of $G \approx 0.44~\si{TeV/m}$.
The mean beam energy is $20.4~\si{MeV}$, yielding a mean acceleration gradient of $G \approx 0.26~\si{TeV/m}$.
The phase space and energy spectrum of the accelerated beam are presented in Fig.~\ref{fig:self_injection}(c) and (d), respectively, which show a broad spectrum.
Electrons continuously transit across the inner surface until the laser pulse is sufficiently depleted, allowing the self-injection process to persist as discussed later.
The high-energy component peaks at $30~\si{MeV}$ with a $26\%$ spread, while the low-energy component peaks at $5.4~\si{MeV}$ with a $1\%$ spread.
This self-injection mechanism is fully feasible with currently available high-power laser facilities, offering a promising pathway for experimental demonstration of this proposed CNT forest-based plasma accelerator.

\begin{figure}
	\includegraphics[width=0.4\textwidth]{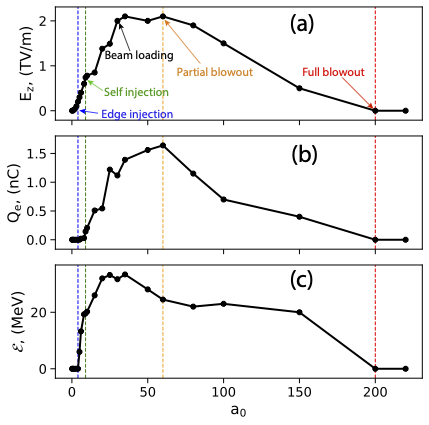}
	\caption{PIC results: Dependence of SPs excitation on normalised laser strength $a_0$. (a)-(c) Peak of acceleration field $E_z$, total charge $Q_e$ and mean energy $\mathcal{E}$ of accelerated electrons after $70~\si{\mu m}$ propagation, respectively. The vertical dashed lines indicate the position where edge (blue) and self-injections (green), partial (orange) and full (red) blowouts begin.}
	\label{fig:SPs_a0}
\end{figure}

Here, it is seen that the key difference between the structured CNT plasma and a uniform gaseous plasma is the presence of the solid inner surface, over which the electrical potential restricts the wall electrons from freely crossing the boundary. 
For an electron on the wall to overcome this potential barrier, it must have sufficient transverse momentum.
Wall electrons gain greater momentum in stronger laser fields, increasing the probability of crossing the inner surface and becoming trapped in the leaky field.
Therefore, for a specific target configuration, the parameter $a_0$ determines both the excitation of SPs and the probability of wall electrons transitioning across the surface for injection.

A series of simulations were performed with $a_0$ values ranging from $0.1$ to $220$. The effects on the amplitude of the acceleration field ($E_z$), the total charge of the injected electron beam ($Q_e$), and the mean energy ($\mathcal{E}$) of the accelerated electron beam are presented in Fig.\ref{fig:SPs_a0} (a)-(c), respectively.
As $a_0$ increases, the SPs become stronger, resulting in electron beams with higher charge that can be accelerated to greater energies.
Edge injection occurs for $4<a_0<9$, while self-injection begins at $a_0\geq 9$, where a significantly larger number of electrons are injected into the leaky field.
At $a_0>30$, the beam loading effect becomes significant, reducing the effective acceleration field and energy gain rate.
At low values of $a_0 < 4$, wall electrons cannot overcome the inner surface potential barrier, preventing any electron injection into the central vacuum channel.
When $a_0<2$, SP becomes non-relativistic.
Simulations also reveal critical thresholds for bubble wakefield excitation and the blowout phenomenon. In a moderately high laser pulse range ($30 < a_0 < 50$), all the wall electrons gain sufficient transverse momentum to freely cross the inner surface. This results in the formation of plasma bubbles, similar to those in uniform plasmas, but with significantly enhanced field amplitudes that exceed the coherent field limit due to the mitigation of wave breaking.
As the laser strength increases further ($a_0 > 50$), wall electrons begin to acquire enough outward momentum from ponderomotive scattering to cross the outer surface of the forest target of finite wall thickness and become ejected from the target—a phenomenon referred to as the blowout effect here. In the range $50 < a_0 < 200$, partial blowout occurs where a fraction of the wall electrons are ejected from the outer wall, leading to a reduction in the acceleration field ($E_z$), total injected charge ($Q_e$), and beam energy ($\mathcal{E}$).
Full blowout occurs at extremely high laser strengths ($a_0 > 200$), where all wall electrons are permanently ejected from the outer surface. Consequently, no sustained acceleration field is formed within the vacuum channel, and particle acceleration ceases entirely.
These thresholds for $a_0$ depend on the balance between laser pulse intensity and target geometry, varying with different target and laser configurations. This interplay highlights the importance of precise tuning to optimise the acceleration process while avoiding undesirable blowout effects.

\begin{figure*}
	\centering
	\includegraphics[width=0.9\textwidth]{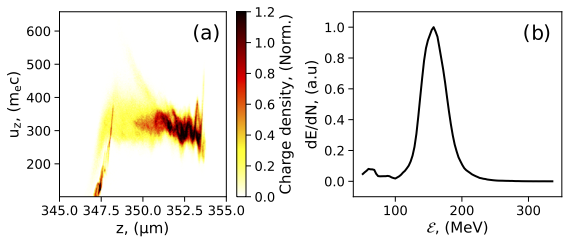}
	\caption{PIC Results:
(a) Phase space and (b) energy spectrum of an electron beam self-injected and accelerated in a $354~\si{\mu m}$-long CNT forest target. The laser and target parameters are identical to those in Fig.~\ref{fig:self_injection}, except for an initial $a_0=30$.}
	\label{fig:cascade}
\end{figure*}

The robustness of SP excitation enables efficient SP-based acceleration in a cascading manner. This approach is highly advantageous for high-power laser-driven experiments, where both beam charge and energy can be significantly increased. Additionally, the energy spectrum of the accelerated electrons can be substantially improved.  
A high-intensity laser pulse can initially drive strong edge injection and self-injection. As the laser energy depletes and the laser strength ($a_0$) decreases, the self-injection process gradually ceases. However, the leaky field remains well-structured and stable, ensuring that the self-trapped electrons continue to be accelerated. This leads to a narrower energy spectrum for the accelerated beam.  
For example, a laser pulse with an initial strength of $a_0 = 30$ can trap an electron beam of charge $2.15~\si{nC}$ in the leaky field, which is accelerated to $180~\si{MeV}$ after $354~\si{\mu m}$ propagation as shown in Fig.~\ref{fig:cascade}. The energy spectrum has a $20\%$ spread and can extend up to $340~\si{MeV}$. This indicates an average acceleration gradient of $G_{\text{AVG}}\approx 0.51~\si{TeV/m}$ and a peak of $G_{\text{PEAK}} \approx 0.96~\si{TeV/m}$.

\section{Conclusion and summary}

In this paper, we propose that a TW optical laser pulse can effectively drive the excitation of strong SPs on the $\mu m$-scale cylindrical surface nanostructured inside a CNT forest.
These SPs feature an EM component that leaks into the central vacuum channel, providing acceleration fields on the order of sub-$\si{TV/m}$. These fields are suitable for accelerating both relativistic negatively and positively charged particles.
We present two distinct electron injection schemes, each with unique features that enhance the experimental feasibility of this approach, paving the way for practical implementations.

In line with recent advances in both laser-driven solid-state and plasma-based accelerators, this paper presents a novel concept for accelerating both positively and negatively charged particles by exciting non-evanescent leaky EM fields of SP modes through the ponderomotive scattering of a high-intensity optical laser pulse passing paraxially through a nanostructured CNT forest.
Compared to conventional conductive materials, such as metals and semiconductors, CNTs can offer significantly higher plasmon gain due to their unique electronic structure and exceptional field confinement. This enables stronger light-matter interactions, which can substantially improve energy efficiency in experimental setups.
Unlike traditional solid-state accelerators, such as crystal-based accelerators or dielectric laser accelerators (DLAs), the CNT-based accelerator leverages plasma dynamics, promising exceptionally high acceleration gradients of up to hundreds of $\si{TV/m}$ in principle and avoinding the limitation from the channel size of natural crystals.
This approach can generate lepton beams in the $\si{MeV}$-level on $\si{\mu m}$ scales or $\si{PeV}$-level on meter scales, offering immense potential for future high-energy applications.
In contrast to gaseous plasma-based accelerators, the CNT-based method is solid-state, capable of providing much higher acceleration fields with enhanced mechanical stability, making it a promising alternative for next-generation acceleration technologies.

The experiments can be designed with the laser facilities of $\si{TW}$ to $\si{PW}$ power with $\si{\mu m}$-scale beam size~\cite{Mourou2019ext}. 
Recent advances in modern nanofabrication techniques have enabled the production of structured CNT forests with specifications that can meet the experimental requirements~\cite{Wang:2024aa}.
This unique combination makes CNT-based particle accelerators a promising solution for pushing the energy frontier in scientific discovery and also poised to significantly enhance our precision in manipulating matter for potential applications in cutting-edge fields like high-energy plasmonics, ion channelling and radiation generation.

\begin{acknowledgements}
Javier Resta-López acknowledges support by the Generalitat Valenciana under grant agreement CIDEGENT/2019/058.
This work made use of the facilities of the N8 Centre of Excellence in Computationally Intensive Research (N8 CIR) provided and funded by the N8 research partnership and EPSRC (Grant No. EP/T022167/1). The Centre is coordinated by the Universities of Durham, Manchester and York.
\end{acknowledgements}

\section*{Data availability}
The datasets used and/or analysed during the current study are available from the corresponding author on reasonable request.

\bibliography{lcnt.bib}

\end{document}